\documentclass[12pt]{article}
\usepackage{graphicx}
\usepackage{amssymb}
\renewcommand{\theequation}{\thesection.\arabic{equation}}

\newlength{\extraspace}
\newlength{\extraspaces}
\newcommand{\be}{\begin{equation}
\addtolength{\abovedisplayskip}{\extraspaces}
\addtolength{\belowdisplayskip}{\extraspaces}
\addtolength{\abovedisplayshortskip}{\extraspace}
\addtolength{\belowdisplayshortskip}{\extraspace}}
\newcommand{\ee}{\end{equation}}

\newcommand{\ba}{\begin{eqnarray}
\addtolength{\abovedisplayskip}{\extraspaces}
\addtolength{\belowdisplayskip}{\extraspaces}
\addtolength{\abovedisplayshortskip}{\extraspace}
\addtolength{\belowdisplayshortskip}{\extraspace}}
\newcommand{\ea}{\end{eqnarray}}

\newcommand{\newsection}[1]{
\vspace{12mm}
\pagebreak[3]
\addtocounter{section}{1}
\setcounter{equation}{0}
\setcounter{subsection}{0}
\setcounter{footnote}{0}
\noindent{\bf \thesection. #1}
\nopagebreak
\medskip
\nopagebreak}

\newcounter{saveeqn}
\newcommand{\alpheqn}{\setcounter{saveeqn}{\value{equation}}%
 \stepcounter{saveeqn}\setcounter{equation}{0}%
 \renewcommand{\theequation}
     {\mbox{\thesection.\arabic{saveeqn}\alph{equation}}}}
\newcommand{\reseteqn}{\setcounter{equation}{\value{saveeqn}}%
  \renewcommand{\theequation}{\thesection.\arabic{equation}}}

\newcommand{\dif}{\mathrm{d}}
\newcommand{\me}{\mathrm{e}}

\begin{document}
\addtolength{\baselineskip}{1.5mm}

\thispagestyle{empty}
\begin{flushright}
hep-th/0511007\\
\end{flushright}
\vbox{}
\vspace{2cm}

\begin{center}
{\LARGE{Accelerating black diholes\\[2mm]
and static black di-rings
        }}\\[16mm]
{Edward Teo}
\\[6mm]
{\it The Abdus Salam International Centre for Theoretical Physics,\\[1mm]
Strada Costiera 11, 34014 Trieste, Italy}\\[5mm]
{\it Department of Physics,
National University of Singapore, 
Singapore 119260}\\[15mm]

\end{center}
\vspace{2cm}

\centerline{\bf Abstract}
\bigskip 
\noindent 
We show how a recently discovered black ring solution with a 
rotating 2-sphere can be turned into two new solutions of 
Einstein--Maxwell--dilaton theory. The first is a four-dimensional
solution describing a pair of oppositely charged, extremal black 
holes---known as a black dihole---undergoing uniform acceleration. 
The second is a five-dimensional solution describing a pair of
concentric, static extremal black rings carrying opposite dipole 
charges---a so-called black di-ring. The properties of both solutions, 
which turn out to be formally very similar, are analyzed in detail. 
We also present, in an appendix, an accelerating version of the 
Zipoy--Voorhees solution in four-dimensional Einstein gravity.


\newpage

\newsection{Introduction}

The complexity of the Einstein field equations, both
in four and especially in higher dimensions, means that 
exact solutions describing physically
interesting space-times are not easy to come by. 
Instead of solving the field equations directly,
one of the more fruitful ways of obtaining new solutions
has been to generate them starting from known solutions.
There is, by now, a large catalogue of `tricks' developed 
to do precisely this, although it is not always clear {\it a priori\/}
that one can obtain physically meaningful solutions by a given 
solution-generating technique.

One of the most spectacular examples of a physically interesting 
solution obtained in this 
way in recent years, is the five-dimensional rotating black ring 
solution of Emparan and Reall \cite{EmparanReall2}. This black ring
is so-called because it has an event horizon of topology $S^1\times S^2$,
and it is the first and only known example of an asymptotically flat,
regular vacuum solution with a horizon of non-spherical topology.
It rotates in the $S^1$ ring direction, creating the necessary
centrifugal force to balance its gravitational self-attraction.
This solution was obtained using the direct relationship between 
five-dimensional Einstein gravity and its four-dimensional
Kaluza--Klein reduction. The starting point was the previously
known four-dimensional 
electrically charged C-metric solution in Kaluza--Klein theory,
describing a Kaluza--Klein black hole undergoing uniform
acceleration. Emparan and Reall lifted this solution to 
five dimensions with appropriate analytic continuations, and then
showed that it can be reinterpreted as a rotating black ring. 

In hindsight, it is possible to understand why there should at all be
a relationship between four-dimensional accelerating black holes
and five-dimensional black rings. This is most clearly seen from
the so-called rod structures of both space-times \cite{EmparanReall}. 
The rod structure of an accelerating space-time would necessarily 
contain a semi-infinite rod for the time coordinate, corresponding to 
the acceleration horizon. Such a rod,
when analytically continued to a space-like coordinate, has the
different interpretation as an axis of rotation. It turns
out that the procedure employed by Emparan and Reall has the effect
of turning the acceleration
horizon into a rotational axis for the new fifth coordinate (while
leaving the rest of the rod structure essentially unchanged). This 
results in the four-dimensional black hole turning into a
five-dimensional black ring around the new axis.

This relationship between the two different space-times has a few 
interesting implications. For example, consider taking the limit of the 
black ring in which the $S^1$ ring radius is sent to infinity.
Note that  
increasing the ring radius corresponds, in the four-dimensional picture,
to increasing the distance between the black hole and the acceleration 
horizon. The limit in which the acceleration horizon is pushed to infinity
is, in fact, equivalent to taking
the zero-acceleration limit of the Kaluza--Klein C-metric.
Thus, we see that there is a formal equivalence 
between taking the infinite-ring radius limit in five dimensions,
and taking the zero-acceleration limit in four dimensions.

Recently, a rotating black ring different from the Emparan--Reall
black ring was discovered by Figueras 
\cite{Figueras}.\footnote{This solution was also claimed to have been
discovered in \cite{Mishima}, although in different and indeed much
more complicated coordinates.}
The difference lies in the fact that the rotation is now in the
azimuthal direction of the $S^2$, rather than in the $S^1$ direction.
Because of this, the self-attraction of the ring is not balanced
by any centrifugal force, and there are necessarily conical singularities
inside the ring to prevent it from collapsing. For the record, this 
black ring was discovered by ``educated guesswork'' \cite{Figueras}, 
from the limits it was expected to reproduce.

In the light of the above-described relationship between five-dimensional black rings 
and four-dimensional accelerating black holes, it is natural to wonder
what this new black ring solution would correspond to in four dimensions.
An important clue comes from taking the infinite-radius limit of it, in which
one obtains a Kerr black hole extended along the (straight) fifth 
direction. Upon taking the appropriate analytic continuations
interchanging the time and fifth coordinates, this becomes the
Euclideanized version of the Kerr black hole with a flat time direction.
The Kaluza--Klein reduction of this space-time would therefore be
the zero-acceleration limit of the, as yet unidentified,
four-dimensional solution.

Now, it turns out that the Kerr black hole has indeed been used before, 
in precisely such a fashion, to generate a new solution of 
Kaluza--Klein theory \cite{Gross}. The resulting solution was found 
to describe a static, magnetic dipole source. This dipole 
source can be interpreted as two oppositely charged, extremal
Kaluza--Klein black holes in static equilibrium, a configuration
known as a black dihole \cite{Emparan1}. Thus, we may conclude that the 
Figueras black ring can be turned, by appropriate analytic continuations 
and Kaluza--Klein reduction, into a new four-dimensional
solution describing an {\it accelerating\/} black dihole.

We shall see in Sec.~2 that this interpretation is indeed correct.
In fact, we would be able to generalize this Kaluza--Klein
solution to one of Einstein--Maxwell--dilaton theory with arbitrary
exponential dilaton coupling. The properties of the accelerating dihole 
solution are then studied in detail. We first show how various limits
of it can be taken, to obtain other known solutions such
as the extremal dilatonic C-metric. It is then shown that there are 
necessarily conical singularities along the axis of symmetry, which are 
responsible for the acceleration of the black holes, as well as for
keeping them apart. This solution is also generalized to 
include a background magnetic field, although it is found that this 
magnetic field is unable to fully replace the roles of the conical 
singularities.

There is another solution that can be generated from the 
Figueras black ring, by applying a higher-dimensional analogue of
the procedure used in \cite{Gross}. Specifically, the black ring
is Euclideanized (with $t\rightarrow ix^6$), and a flat time direction 
is added. The resulting six-dimensional vacuum solution can then
be dimensionally reduced along $x^6$ to give a new solution of 
five-dimensional Kaluza--Klein theory. Recall that when this procedure is 
applied to the five-dimensional Myers--Perry black hole, 
a static, extremal magnetic black ring results \cite{Dowker2,Emparan2}.
Since diametrically opposite points of this ring carry opposite
charges, the ring has a zero net charge; for this reason, it is also
known as a dipole ring. Now, it turns out that this new solution
describes a pair of concentric, extremal dipole rings. Furthermore, they
carry opposite charges, in the sense that points on either ring
with the same $S^1$ coordinate have opposite charges. In analogy 
with the case of black diholes, we shall call such a double-ring
configuration a `black di-ring'.

This solution is the subject of Sec.~3. In fact, we would be able
to generalize the Kaluza--Klein di-ring to one of five-dimensional
Einstein--Maxwell--dilaton theory with arbitrary exponential 
dilaton coupling.\footnote{Rings in this theory have previously been
considered, e.g., in \cite{Kunduri}, but these have a {\it net\/} electric
charge, and so are different from the ones considered in this paper.} 
It turns out that this di-ring solution is
formally very similar to the accelerating dihole solution, and so
the analysis of its properties would proceed analogously.
Various limits of it are first taken to confirm its interpretation.
It is then shown that conical singularities are necessarily
present in the system to counterbalance the self-attraction of the inner
ring, as well as the mutual attraction of the two rings.
This solution is also generalized to include a background magnetic
field, although it is found that this magnetic field is not able to
fully balance the forces present and remove all the conical singularities
from the system.

The paper concludes with some brief remarks, including possible 
applications to string theory. There is also 
an appendix, in which an accelerating version 
of the Zipoy--Voorhees solution \cite{Zipoy,Voorhees}
in four-dimensional vacuum Einstein gravity is presented. 
This solution was deduced from the fact that a special case
of the Zipoy--Voorhees solution (the so-called Darmois solution)
arises as the coincident limit of the black dihole solution in
pure Einstein--Maxwell theory.

\newsection{Accelerating black dihole solution}

Our starting point is the metric for the black ring with a rotating 
2-sphere, as found by Figueras \cite{Figueras}:
\ba
\label{BR}
\dif s^2&=&-\frac{H(y,x)}{H(x,y)}\left[\dif t-\frac{2maAy(1-x^2)}
{H(y,x)}\,\dif\varphi\right]^2\nonumber\\
&&+\frac{1}{A^2(x-y)^2}\,H(x,y)\Bigg[-\frac{\dif y^2}{(1-y^2)F(y)}
-\frac{(1-y^2)F(x)}{H(x,y)}\,\dif\psi^2\nonumber\\
&&\phantom{+\frac{1}{A^2(x-y)^2}\,H(x,y)\Bigg[}+\frac{\dif x^2}{(1-x^2)F(x)}
+\frac{(1-x^2)F(y)}{H(y,x)}\,\dif\varphi^2\Bigg]\,,
\ea
where
\be
F(\xi)=1+2mA\xi+(aA\xi)^2,\qquad 
H(\xi_1,\xi_2)=1+2mA\xi_1+(aA\xi_1\xi_2)^2,
\ee
and $m$, $a$ and $A$ are positive constants.
The angular coordinate $\psi$ parametrizes the $S^1$ direction of the
ring, while $(x,\varphi)$ parametrize the rotating $S^2$. Further 
details on this solution may be found in \cite{Figueras}.

We now perform the analytic continuation $t\rightarrow ix^5$, 
$\psi\rightarrow it$, $a\rightarrow ia$, and dimensionally 
reduce along the fifth direction $x^5$. The result is a solution
to Kaluza--Klein theory with the following action:
\be
\label{KK}
\frac{1}{16\pi G}\int\dif^4x\,\sqrt{-g}\left(R-\frac{1}{2}(\partial\phi)^2
-\frac{1}{4}\me^{-\sqrt{3}\phi}F^2\right),
\ee
where $F_{ab}\equiv\partial_aA_b-\partial_bA_a$. The four-dimensional 
metric is 
\alpheqn\ba
\label{KK_metric}
\dif s^2&=&\frac{1}{A^2(x-y)^2}\Bigg[
\left(\frac{H(y,x)}{H(x,y)}\right)^{\frac{1}{2}}
(1-y^2)F(x)\,\dif t^2+\left(\frac{H(y,x)}{H(x,y)}\right)^{-\frac{1}{2}}(1-x^2)F(y)\,\dif\varphi^2\nonumber\\
&&\phantom{\frac{1}{A^2(x-y)^2}\Bigg[}+\frac{(H(x,y)H(y,x))^{\frac{1}{2}}}{K_0^2}\Bigg(-\frac{\dif y^2}{(1-y^2)F(y)}
+\frac{\dif x^2}{(1-x^2)F(x)}\Bigg)\Bigg]\,,
\ea
while the gauge potential and dilaton are, respectively,
\ba
\label{KK_A}
A_\varphi&=&\frac{2maAy(1-x^2)}{H(y,x)}\,,\\
\phi&=&-\frac{\sqrt{3}}{2}\ln\left(\frac{H(y,x)}{H(x,y)}\right),
\ea\reseteqn
where, now,
\be
\label{FH}
F(\xi)=1+2mA\xi-(aA\xi)^2,\qquad 
H(\xi_1,\xi_2)=1+2mA\xi_1-(aA\xi_1\xi_2)^2.
\ee
Note that we have introduced a constant $K_0$ into (\ref{KK_metric}),
whose significance would be discussed below. 
This solution is manifestly static and axisymmetric about the $\varphi$-axis. 
As can be seen from (\ref{KK_A}), it carries a pure magnetic charge.

Kaluza--Klein theory is, in fact, a special case of a more general
class of Einstein--Maxwell--dilaton theories with the action
\be
\label{4Daction}
\frac{1}{16\pi G}\int\dif^4x\,\sqrt{-g}\left(R-\frac{1}{2}(\partial\phi)^2
-\frac{1}{4}\me^{-\alpha\phi}F^2\right),
\ee
where $\alpha$ is a non-negative parameter known as the dilaton coupling.
It would prove to be convenient to introduce the new parameter 
$N$ given by
\be
\label{N}
N=\frac{4}{1+\alpha^2}\,,\qquad 0<N\leq4\,.
\ee
The action (\ref{KK}) is recovered when $N=1$, while other integer
values of $N$ also correspond to well-known cases. One would naturally 
like to keep the dilaton coupling general as far as possible.

Fortunately, there is a systematic procedure known \cite{Liang,EmparanTeo}
to turn a static, axisymmetric, magnetic solution of (\ref{KK})
into one of (\ref{4Daction}), valid for general dilaton coupling. When 
applied to (2.4), 
the resulting solution is
\alpheqn\ba
\label{metric}
\dif s^2&=&\frac{1}{A^2(x-y)^2}\Bigg[
\left(\frac{H(y,x)}{H(x,y)}\right)^{\frac{N}{2}}
(1-y^2)F(x)\,\dif t^2+\left(\frac{H(y,x)}{H(x,y)}\right)^{-\frac{N}{2}}(1-x^2)F(y)\,\dif\varphi^2\nonumber\\
&&\phantom{\frac{1}{A^2(x-y)^2}\Bigg[}+\frac{(H(x,y)H(y,x))^{\frac{N}{2}}}{K_0^2\,G(x,y)^{N-1}}\,\Bigg(-\frac{\dif y^2}{(1-y^2)F(y)}
+\frac{\dif x^2}{(1-x^2)F(x)}\Bigg)\Bigg]\,,\\
A_\varphi&=&\frac{2\sqrt{N}maAy(1-x^2)}{H(y,x)}\,,\\
\phi&=&-\frac{\alpha N}{2}\ln\left(\frac{H(y,x)}{H(x,y)}\right),
\ea\reseteqn
where $F(\xi)$ and $H(\xi_1,\xi_2)$ are as in (\ref{FH}), and the new
function $G(x,y)$ is given by
\be
\label{G}
G(x,y)=(1+mA(x+y)-a^2A^2xy)^2-(m^2+a^2)A^2(1-xy)^2.
\ee
It is to this general solution that we shall instead direct our attention. 
We will see that it describes a black dihole undergoing uniform 
acceleration, and the analysis of its properties will closely follow 
both that of the C-metric describing a single accelerating black hole 
(see, e.g., \cite{Dowker1,Hong}), 
and the non-accelerating dihole solution \cite{Emparan1}.

Let us write the two roots of $F(y)$ as $-\frac{1}{r_\pm A}$, where
\be
r_\pm\equiv m\pm\sqrt{m^2+a^2}\,.
\ee
Furthermore, we assume that $0<r_+A<1$. Then 
to ensure that the metric (\ref{metric}) has the correct space-time 
signature, the $(x,y)$ coordinates have to take the range
\be
\label{xy_range}
-1\leq x\leq1\,,\qquad -\frac{1}{r_+A}\leq y\leq -1\,.
\ee
The curvature invariants of the metric show that asymptotic 
infinity is at $x=y=-1$, while the only curvature singularities 
in this range lie at the two points $y=-\frac{1}{r_+A}$, $x=\pm1$.
 
We now construct the rod structure \cite{EmparanReall} of this metric, 
which would help to reveal its physical significance. Note that the metric
component $g_{tt}$ vanishes at the two points $y=-\frac{1}{r_+A}$, 
$x=\pm1$, which indicates that these are the locations of two extremal
black holes.\footnote{Recall that extremal black holes have the 
universal property that their corresponding rods have zero length (see,
e.g., \cite{EmparanTeo}).} This is consistent with the above-mentioned 
fact that the curvature is infinite at the same two points.
$g_{tt}$ also vanishes along the semi-infinite line $y=-1$, 
which may be identified as an acceleration horizon \cite{EmparanReall}. 
Thus, we arrive at the rod structure in Fig.~1, which clearly corresponds 
to two extremal black holes undergoing acceleration.
The part of the symmetry axis between the two black holes is given by 
$y=-\frac{1}{r_+A}$; that between the first black hole (as labelled in
Fig.~1) and the acceleration horizon is $x=1$, while that joining the 
second black hole to infinity is $x=-1$. 

This physical interpretation of the solution can be confirmed by taking
various limits of it. A familiar one is the zero-acceleration limit,
in which the acceleration horizon is effectively pushed to infinity. 
This is achieved by performing the coordinate transformation
\be
\label{zero_acc}
t=A\tilde t\,,\qquad x=\cos\theta\,,\qquad y=-\frac{1}{rA}\,,
\ee
and then taking $A\rightarrow0$. In this limit, the solution 
(2.8) 
reduces to 
\alpheqn\ba
\label{dihole_metric}
\dif s^2&=&\left(\frac{\triangle+a^2\sin^2\theta}{\Sigma}\right)^{\frac{N}{2}}
\left[-\dif\tilde t^2+\frac{\Sigma^{N}}{K_0^2\,(\triangle+(m^2+a^2)\sin^2\theta)^{N-1}}\Bigg(\frac{\dif r^2}{\triangle}
+\dif\theta^2\Bigg)\right]\nonumber\\
&&\qquad
+\left(\frac{\triangle+a^2\sin^2\theta}{\Sigma}\right)^{-\frac{N}{2}}
\triangle\sin^2\theta\,\dif\varphi^2,\\
\label{dihole_gauge}
A_\varphi&=&-\frac{2\sqrt{N}mra\sin^2\theta}{\triangle+a^2\sin^2\theta}\,,\\
\label{dihole_phi}
\phi&=&-\frac{\alpha N}{2}\ln\left(\frac{\triangle+a^2\sin^2\theta}
{\Sigma}\right),
\ea\reseteqn
where
\be
\label{triangle}
\triangle=r^2-2mr-a^2,\qquad\Sigma=r^2-a^2\cos^2\theta\,.
\ee
This is but the solution for a dilatonic black dihole \cite{Davidson,Galtsov}. 
It not only confirms the 
presence of two extremal magnetic black holes in the rod structure 
of Fig.~1, it also proves that they have opposite charges.
Furthermore, it is known from the dihole solution that the parameters $m$ 
and $a$ are related to the mass/charge of the black holes and the 
separation between them, respectively, and this would continue to
be true in the general solution.

\begin{figure}[t]
\begin{center}
\includegraphics{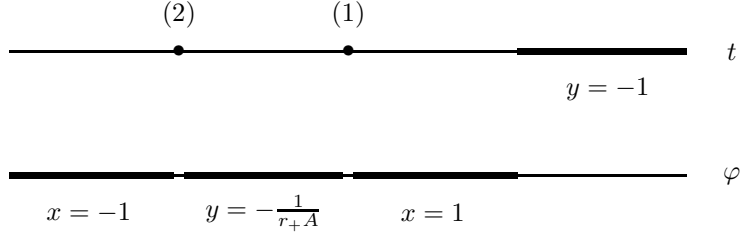}
\caption{The rod structure of the accelerating dihole solution.}
\end{center}
\end{figure}

Another limit one could consider is to push the second
black hole of Fig.~1 to infinity, while leaving the first black hole behind.
One would then expect to obtain the usual
C-metric describing a single extremal black hole undergoing 
uniform acceleration.
This limit is achieved if we set $A=\tilde A/(1+a\tilde A)$,
perform the coordinate transformation
\be
\label{C_limit}
x=1-\frac{(1-\tilde x)(1+m\tilde A\tilde y)}{a\tilde A(\tilde x-\tilde y)}\,,
\qquad
y=-1+\frac{(1+\tilde y)(1+m\tilde A\tilde x)}{a\tilde A(\tilde x-\tilde y)}\,,
\ee
and then take $a\rightarrow\infty$. 
The solution (2.8) 
becomes, for an appropriate choice of $K_0$,
\alpheqn\ba
\dif s^2&=&\frac{1}{\tilde A^2(\tilde x-\tilde y)^2}\Bigg[
\left(\frac{F(\tilde y)}{F(\tilde x)}\right)^{\frac{N}{2}}
(1-\tilde y^2)F(\tilde x)^2\,\dif t^2+\left(\frac{F(\tilde y)}{F(\tilde x)}\right)^{-\frac{N}{2}}(1-\tilde x^2)F(\tilde y)^2\,\dif\varphi^2\nonumber\\
&&\phantom{\frac{1}{\tilde A^2(\tilde x-\tilde y)^2}\Bigg[}+(F(\tilde x)F(\tilde y))^{\frac{4-N}{2}}\,\Bigg(-\frac{\dif\tilde y^2}{(1-\tilde y^2)F(\tilde y)^2}
+\frac{\dif\tilde x^2}{(1-\tilde x^2)F(\tilde x)^2}\Bigg)\Bigg]\,,\\
A_\varphi&=&-\sqrt{N}m(1-\tilde x)\,,\\
\phi&=&-\frac{\alpha N}{2}\ln\left(\frac{F(\tilde y)}{F(\tilde x)}\right),
\ea\reseteqn
where $F(\tilde\xi)=1+m\tilde A\tilde\xi$. Indeed, it can be seen that 
this is just the dilatonic generalization of the extremal magnetic C-metric 
\cite{Dowker1}, in the form written in \cite{Hong}.

It is also possible to zoom in to each of the black holes, without
the need to send the other black hole or the acceleration horizon 
to infinity. To zoom in to the first black hole, we perform the
coordinate transformation
\alpheqn\ba
t&=&\frac{A\tilde t}{\sqrt{(1-r_+A)(1+r_-A)}}\,,\\
\label{bh1_x}
x&=&1-\frac{1+r_+A}{2\sqrt{m^2+a^2}}\sqrt{\frac{1+r_-A}{1-r_+A}}\,r(1-\cos\theta)\,,\\
y&=&-\frac{1}{r_+A}\left[1-\frac{1+r_+A}{2r_+}\sqrt{\frac{1-r_+A}{1+r_-A}}\,r(1+\cos\theta)\right],
\ea\reseteqn
and then take the new coordinate $r$ to be smaller than any other 
length scale present. For an appropriately chosen $K_0$, the solution 
(2.8) 
becomes 
\alpheqn\ba
\label{bh1_metric}
\dif s^2&=&g(\theta)^{\frac{N}{2}}
\left[-\left(\frac{r}{Q}\right)^{\frac{N}{2}}\dif\tilde t^2+\left(\frac{r}{Q}\right)^{-\frac{N}{2}}(\dif r^2
+r^2\dif\theta^2)\right]\nonumber\\
&&\qquad
+\left(\frac{r}{Q}\,g(\theta)\right)^{-\frac{N}{2}}
r^2\sin^2\theta\,\dif\varphi^2,\\
\label{bh1_gauge}
A_\varphi&=&-\frac{aQ\sqrt{N}}{\sqrt{m^2+a^2}}\,
\sqrt{\frac{1+r_-A}{1-r_+A}}\,
\frac{1-\cos\theta}{g(\theta)}\,,\\
\label{bh1_phi}
\phi&=&-\frac{\alpha N}{2}\ln\left(\frac{r}{Q}\,g(\theta)\right),
\ea\reseteqn
where we have defined 
\alpheqn\ba
\label{Q}
Q&=&\frac{mr_+}{\sqrt{m^2+a^2}}\,\sqrt{\frac{1+r_-A}{1-r_+A}}\,,\\
\label{g}
g(\theta)&=&\frac{1}{2}\left[1+\cos\theta+\frac{a^2}{m^2+a^2}\,\frac{1+r_-A}{1-r_+A}\,(1-\cos\theta)\right].
\ea\reseteqn
Note that if the factor $g(\theta)$ were replaced by 1, the metric
(\ref{bh1_metric}) would be precisely that near the horizon of 
an extremal dilaton black hole (with the horizon located at $r=0$). 
In actual fact, this horizon is
distorted away from spherical symmetry by a non-trivial $g(\theta)$, 
and this may be attributed to the effect of the other black hole as
well as the acceleration.
Furthermore, a calculation of the curvature invariants reveals that
this horizon is regular only in the non-dilatonic case $N=4$. This
is consistent with the fact that extremal dilaton black holes have
horizons which are actually null singularities.

Similarly, we may zoom in to the second black hole with 
the coordinate transformation
\alpheqn\ba
t&=&\frac{A\tilde t}{\sqrt{(1+r_+A)(1-r_-A)}}\,,\\
\label{bh2_x}
x&=&-1+\frac{1-r_+A}{2\sqrt{m^2+a^2}}\sqrt{\frac{1-r_-A}{1+r_+A}}\,r(1-\cos\theta)\,,\\
y&=&-\frac{1}{r_+A}\left[1-\frac{1-r_+A}{2r_+}\sqrt{\frac{1+r_+A}{1-r_-A}}\,r(1+\cos\theta)\right].
\ea\reseteqn
The resulting solution is again given by 
(2.18) and (2.19), 
but with the replacement $A\rightarrow-A$.
This reflects the asymmetry of the two black holes with respect
to the location of the acceleration horizon; in other words, they are 
affected differently by the acceleration. But the above remarks for the 
first black hole still apply.

This distortion of the two black-hole horizons is
related to the fact that there are conical singularities 
along the symmetry axis attached to the black holes.
If we take the angle $\varphi$ to have the usual periodicity $2\pi$, 
then the deficit angles along the three different parts of the axis
are
\alpheqn\ba
\label{delta1}
\delta_{(x=1)}&=&2\pi\,\Big[1-(1+2mA-a^2A^2)^{\frac{N}{2}}\,K_0\Big]\,,\\
\label{delta2}
\delta_{(x=-1)}&=&2\pi\,\Big[1-(1-2mA-a^2A^2)^{\frac{N}{2}}\,K_0\Big]\,,\\
\delta_{\big(y=-\frac{1}{r_+A}\big)}&=&2\pi\,\bigg\{1-\bigg[\bigg(1+\frac{m^2}{a^2}\bigg)(1-r_+^2A^2)\bigg]^{\frac{N}{2}}\,K_0\bigg\}\,.
\ea\reseteqn
In general, it is possible to remove only one of the three conical 
singularities with an appropriate choice of the constant $K_0$. 
For example, if we remove the conical singularity along 
$x=1$ with the choice 
$K_0=(1+2mA-a^2A^2)^{-\frac{N}{2}}$, then there
is a positive deficit angle along $x=-1$. This can be interpreted as a
semi-infinite cosmic string pulling on the dihole pair. 
Alternatively, we can remove the conical singularity along $x=-1$
with the choice $K_0=(1-2mA-a^2A^2)^{-\frac{N}{2}}$,
resulting in a {\it negative\/} deficit angle along $x=1$. This
can be interpreted as a strut pushing on the dihole pair.
The strut actually continues past the acceleration horizon and joins
up with a `mirror' dihole pair on the other side of it, although 
a change of coordinates is needed to see this \cite{Bonnor2}.

In both the preceding cases, there is also in general a conical 
singularity along the axis $y=-\frac{1}{r_+A}$ between the pair of 
black holes. When 
$K_0=(1-2mA-a^2A^2)^{-\frac{N}{2}}$,
it can be checked that the deficit angle along $y=-\frac{1}{r_+A}$ is
always negative. Thus, in addition to the strut along $x=1$ pushing on 
the first black hole, there is another strut between 
the two black holes keeping them apart.
On the other hand, when $K_0=(1+2mA-a^2A^2)^{-\frac{N}{2}}$, the deficit 
angle along $y=-\frac{1}{r_+A}$ can take either sign.
A particularly interesting situation is when it vanishes, which occurs
when
\be
1+\frac{m^2}{a^2}=\frac{1+r_-A}{1-r_+A}\,.
\ee
In this situation, the {\it only\/} conical singularity in the 
space-time is the cosmic string along $x=-1$ pulling on the second black hole.
The first black hole is accelerated
along in the same direction by virtue of the attraction between the
two black holes. Note that in this case, the distortion factor (\ref{g})
for the first black hole's horizon is identically one; since there are 
no conical singularities attached to this black hole, its horizon is 
perfectly spherically symmetric.

Now, recall that it is possible to immerse the dilatonic C-metric 
\cite{Dowker1} or dihole solution \cite{Emparan1} in a background 
magnetic field, by means of a Harrison-type transformation 
\cite{Dowker1}. 
By adjusting its field strength appropriately, such a magnetic field could 
replace the role of the conical singularities in accelerating the 
black holes or keeping them apart. It is similarly possible to
immerse the accelerating dihole solution (2.8) 
in a background magnetic field. The resulting solution is
\alpheqn\ba
\label{mag_metric}
\dif s^2&=&\frac{1}{A^2(x-y)^2}\Bigg[
\left(\frac{H(y,x)\Lambda}{H(x,y)}\right)^{\frac{N}{2}}
(1-y^2)F(x)\,\dif t^2+\left(\frac{H(y,x)\Lambda}{H(x,y)}\right)^{-\frac{N}{2}}(1-x^2)F(y)\,\dif\varphi^2\nonumber\\
&&\phantom{\frac{1}{A^2(x-y)^2}\Bigg[}+\frac{(H(x,y)H(y,x)\Lambda)^{\frac{N}{2}}}{K_0^2\,G(x,y)^{N-1}}\,\Bigg(-\frac{\dif y^2}{(1-y^2)F(y)}
+\frac{\dif x^2}{(1-x^2)F(x)}\Bigg)\Bigg]\,,\\
\label{mag_gauge}
A_\varphi&=&\frac{2\sqrt{N}maAy(1-x^2)}{H(y,x)\Lambda}
+\frac{B(1-x^2)}{H(y,x)^2\Lambda}
\left((2maAy)^2(1-x^2)+\frac{F(y)H(x,y)^2}{A^2(x-y)^2}\right),\\
\label{mag_phi}
\phi&=&-\frac{\alpha N}{2}\ln\left(\frac{H(y,x)\Lambda}{H(x,y)}\right),
\ea\reseteqn
where the function $\Lambda$ is given by
\be
\label{Lambda}
\Lambda=\left(1+\frac{2B}{\sqrt{N}}\,\frac{maAy(1-x^2)}{H(y,x)}\right)^2
+\frac{B^2}{N}\,\frac{(1-x^2)F(y)}{A^2(x-y)^2}
\left(\frac{H(x,y)}{H(y,x)}\right)^2,
\ee
and $B$ is a new parameter governing the strength of the background 
magnetic field.

The properties of this solution can be analyzed in the usual manner.
For example, we may zoom in to the near-horizon region of either black 
hole using the same 
transformations as above. For the first black hole, we again obtain
the solution (2.18) and (2.19), 
except that $A_\varphi$
acquires the extra factor $1-\frac{2}{\sqrt{N}}\frac{Bmr_+}{a}$, and
$g(\theta)$ is replaced by
\be
\label{g_mag}
g(\theta)=\frac{1}{2}\left[1+\cos\theta+\frac{a^2}{m^2+a^2}\,\frac{1+r_-A}{1-r_+A}\,\bigg(1-\frac{2}{\sqrt{N}}\,\frac{Bmr_+}{a}\bigg)^2(1-\cos\theta)\right].
\ee
The near-horizon solution for the second black hole is obtained from this by replacing
$A\rightarrow-A$. Thus, the magnetic field does contribute to the 
distortion of the black-hole horizons.

One may also check for the presence of conical singularities in this space-time 
in the usual manner. It turns out that the deficit angles along the 
$x=\pm1$ axes are unchanged from (\ref{delta1},b).
Thus, the above arguments are still valid: there must either be
a cosmic string along $x=-1$ pulling on the dihole pair, or 
a strut along $x=1$ pushing on it. The background magnetic field
cannot be the source of acceleration, unlike in the case of the charged
C-metric. The reason for this is simple:
the dihole pair has a zero net charge, and so its overall motion is
unaffected by the magnetic field. 
On the other hand, the conical deficit
along $y=-\frac{1}{r_+A}$ is now replaced by
\be
\label{B_delta}
\delta_{\big(y=-\frac{1}{r_+A}\big)}=2\pi\,\bigg\{1-\bigg[\bigg(1+\frac{m^2}{a^2}\bigg)(1-r_+^2A^2)\bigg(1-\frac{2}{\sqrt{N}}\,\frac{Bmr_+}{a}\bigg)^{-2}\bigg]^{\frac{N}{2}}\,K_0\bigg\}\,.
\ee
This means that the magnetic field can play a role in balancing the forces
between the two
black holes. In fact, by adjusting $B$ appropriately, it is
always possible to remove the conical singularity along $y=-\frac{1}{r_+A}$,
regardless of the choice of $K_0$. Again, if all the conical singularities
attached to a particular black hole is removed, it can be checked that
the horizon of this black hole becomes perfectly spherically symmetric.

Finally, we note that electric versions of the above solutions may
be obtained by performing the duality transformation:
\be
\phi'=-\phi\,,\qquad F_{ab}'=\frac{\me^{-\alpha\phi}}{2}\,
\varepsilon_{ab}{}^{cd}F_{cd}\,.
\ee
Applying this transformation to the solution (2.23) 
gives the following electric gauge potential: 
\ba
\label{electric}
A_t'&=&\frac{2\sqrt{N}maAx(1-y^2)}{H(x,y)}\left(1-\frac{B}{\sqrt{N}}\frac{ay(1-x^2)}{x-y}\right)^2\nonumber\\
&&-\frac{B}{A^2(x-y)^2}\bigg((1-2xy+x^2)\Big(1+4mAy\Big(1+\frac{Ba}{\sqrt{N}}\Big)\Big)\nonumber\\
&&\phantom{-\frac{B}{A^2(x-y)^2}\Big(}-Ax(1-y^2)\Big(2m+a^2Ax+\frac{2B}{\sqrt{N}}\,ma(1+x^2)\Big)\bigg)\,.
\ea
This dual solution describes an electrically charged dihole accelerating
in a background electric field. Its properties are otherwise very similar
to those of the magnetic solution.

\newsection{Black di-ring solution}

Our starting point is again the black ring metric (\ref{BR}).
We now perform the analytic continuation $t\rightarrow ix^6$, 
$a\rightarrow ia$, to obtain a Euclidean version of this metric. 
A flat time direction is then added, resulting in a six-dimensional
vacuum Einstein metric with Lorentzian signature. When dimensionally
reduced along $x^6$, we obtain a solution to five-dimensional 
Kaluza--Klein theory with the action
\be
\label{5DKK}
\frac{1}{16\pi G}\int\dif^5x\,\sqrt{-g}\left(R-\frac{1}{2}(\partial\phi)^2
-\frac{1}{4}\me^{-2\sqrt{\frac{2}{3}}\phi}F^2\right).
\ee
Explicitly, the solution is
\alpheqn\ba
\label{5DKK_metric}
\dif s^2&=&-\left(\frac{H(y,x)}{H(x,y)}\right)^{\frac{1}{3}}\dif t^2
+\frac{1}{A^2(x-y)^2}\,\left(H(y,x)H(x,y)^2\right)^{\frac{1}{3}}\Bigg[
-\frac{(1-y^2)F(x)}{H(x,y)}\,\dif\psi^2\nonumber\\
&&\qquad+\frac{1}{K_0^2}\left(
-\frac{\dif y^2}{(1-y^2)F(y)}
+\frac{\dif x^2}{(1-x^2)F(x)}\right)
+\frac{(1-x^2)F(y)}{H(y,x)}\,\dif\varphi^2\Bigg]\,,\\
A_\varphi&=&\frac{2maAy(1-x^2)}{H(y,x)}\,,\\
\phi&=&-\sqrt{\frac{2}{3}}\ln\left(\frac{H(y,x)}{H(x,y)}\right),
\ea\reseteqn
where $F(\xi)$ and $H(\xi_1,\xi_2)$ are as in (\ref{FH}).
Note that we have introduced a constant $K_0$ into (\ref{5DKK_metric}),
which will be adjusted appropriately below.

As in four dimensions, the action (\ref{5DKK}) is but a special case
of a more general class of five-dimensional Einstein--Maxwell--dilaton
theories with the action
\be
\label{5D_action}
\frac{1}{16\pi G}\int\dif^5x\,\sqrt{-g}\left(R-\frac{1}{2}(\partial\phi)^2
-\frac{1}{4}\me^{-\alpha\phi}F^2\right).
\ee
It will turn out to be convenient to introduce the new parameter 
\be
\label{5D_N}
N=\frac{12}{4+3\alpha^2}\,,\qquad 0<N\leq3\,,
\ee
so that the Kaluza--Klein case 
corresponds to $N=1$. Again, it is straightforward to generalize
the Kaluza--Klein solution (3.2) 
to other values of the dilaton coupling. The general solution is
\alpheqn\ba
\label{5D_metric}
\dif s^2&=&-\left(\frac{H(y,x)}{H(x,y)}\right)^{\frac{N}{3}}\dif t^2
+\frac{1}{A^2(x-y)^2}\,\left(H(y,x)H(x,y)^2\right)^{\frac{N}{3}}\Bigg[
-\frac{(1-y^2)F(x)}{H(x,y)^N}\,\dif\psi^2\nonumber\\
&&+\frac{1}{K_0^2\,G(x,y)^{N-1}}\left(
-\frac{\dif y^2}{(1-y^2)F(y)}
+\frac{\dif x^2}{(1-x^2)F(x)}\right)
+\frac{(1-x^2)F(y)}{H(y,x)^N}\,\dif\varphi^2\Bigg]\,,\\
A_\varphi&=&\frac{2\sqrt{N}maAy(1-x^2)}{H(y,x)}\,,\\
\phi&=&-\frac{\alpha N}{2}\ln\left(\frac{H(y,x)}{H(x,y)}\right),
\ea\reseteqn
where the function $G(x,y)$ is the same as in (\ref{G}).
Given the formal similarity of this solution with the four-dimensional
accelerating dihole (2.8)---in fact, 
only the metric differs---its physical interpretation
and properties would not be too difficult to deduce. In particular,
the $(x,y)$ coordinates continue to take the range (\ref{xy_range}).

The rod structure for this space-time is depicted in Fig.~2. 
Indeed, the resemblance with that in Fig.~1 is obvious, the only
difference being that the acceleration horizon of the four-dimensional
space-time is now
the semi-infinite axis of the new angular coordinate $\psi$.
Hence the space-time contains two concentric, static extremal 
black rings circling around the $\psi$-axis. This interpretation can
be confirmed by taking various limits of the solution.

\begin{figure}[t]
\begin{center}
\includegraphics{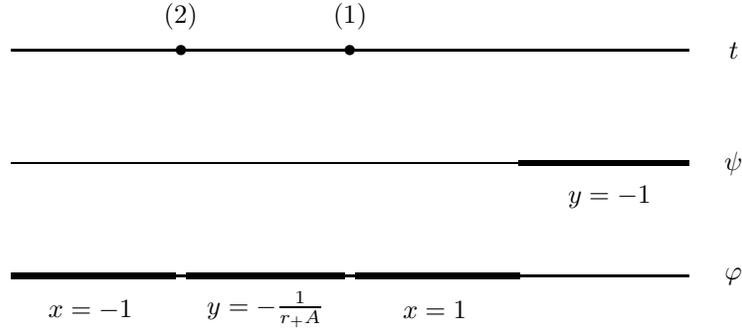}
\caption{The rod structure of the di-ring solution.}
\end{center}
\end{figure}

The simplest limit to take is that of infinite ring radius, which
as explained in the introduction, is formally identical to taking the
zero-acceleration limit in four dimensions. In this case, the 
coordinate transformation is given by (compare (\ref{zero_acc})):
\be
\psi=A\tilde\psi\,,\qquad x=\cos\theta\,,\qquad y=-\frac{1}{rA}\,.
\ee
Upon taking the limit $A\rightarrow0$, we obtain the metric
\ba
\dif s^2&=&\left(\frac{\triangle+a^2\sin^2\theta}{\Sigma}\right)^{\frac{N}{3}}
\left[-\dif t^2+\dif\tilde\psi^2+\frac{\Sigma^{N}}{K_0^2\,(\triangle+(m^2+a^2)\sin^2\theta)^{N-1}}\Bigg(\frac{\dif r^2}{\triangle}
+\dif\theta^2\Bigg)\right]\nonumber\\
&&\qquad
+\left(\frac{\triangle+a^2\sin^2\theta}{\Sigma}\right)^{-\frac{2N}{3}}
\triangle\sin^2\theta\,\dif\varphi^2,
\ea
where $\triangle$ and $\Sigma$ are as in (\ref{triangle}).
The associated gauge field and dilaton are given by (\ref{dihole_gauge}) and
(\ref{dihole_phi}), respectively; of course, this is with the understanding
that $N$ is now defined by (\ref{5D_N}) rather than (\ref{N}).
This solution is in fact a five-dimensional generalization of the 
black dihole solution, whereby the two oppositely charged, extremal magnetic black 
holes are now black strings extended in the $\tilde\psi$ direction. 
These black strings carry a so-called `local charge' \cite{Emparan3}, 
defined by
\be
\label{charge}
{\cal Q}=\frac{1}{4\pi}\int_{S^2}F\,,
\ee
where the $S^2$ encloses a point along the string. The two strings
have opposite local charges, in the sense that when the $S^2$ encloses
points on both strings, the net charge ${\cal Q}$ will be zero.
There will, however, be a residual dipole moment.

This limit shows that the general solution (3.5) 
describes two oppositely charged, extremal magnetic black strings that are
circular, i.e., black rings. Before discussing this configuration,
let us recall what happens in the case of a single black ring.
Such a ring carries a local distribution of charge that continues
to be given by (\ref{charge}). However, if the $S^2$ is enlarged
to enclose diametrically opposite points of the ring, the 
net charge ${\cal Q}$ will be zero. This is due to the opposite
orientations of the individual $S^2$s enclosing the two points in five
dimensions. Thus, the ring as a whole will have a zero net charge, and for 
this reason, it is also known as a dipole ring \cite{Emparan3}.\footnote{However, 
note that the context of the term `dipole' is different from the one 
used in the preceding paragraph.}

Returning to the case of the two oppositely charged black rings, 
we see that there is already a local cancellation of charge 
when the $S^2$ encloses a point from each ring with the same 
$S^1$-coordinate $\psi$. We emphasize that this is even before the
$S^2$ is extended to enclose the other two diametrically opposite points.
When extended, the $S^2$ would enclose two dipoles with {\it opposite\/} 
dipole moments, resulting in a residual quadrupole moment.\footnote{This 
is in fact very similar to 
the situation when the accelerating dihole solution is
extended past the acceleration horizon. There is another dihole pair 
lying in this new region of the space-time, mirroring the original 
dihole pair in the sense that its dipole moment is reversed.
The extended space-time thus contains two dihole pairs oppositely
oriented along a straight line, resulting in a quadrupole moment.}
In analogy with the case of black diholes, we shall refer to such an
oppositely charged double-black-ring configuration as a 
`black di-ring'.

Another limit one can consider to confirm this di-ring interpretation, 
is to send the second black ring of Fig.~2 to infinity. One would 
then expect to recover the solution for a single static, extremal 
dipole ring, first constructed in \cite{Dowker2,Emparan2}. 
This is achieved by setting $A=1/(a+\sqrt{m^2+R^2})$, performing the 
coordinate transformation
\be
x=1-\frac{(\sqrt{m^2+R^2}+m)(1-\tilde x)}{a(\tilde x-\tilde y)}\,,\qquad
y=-1+\frac{(\sqrt{m^2+R^2}-m)(1+\tilde y)}{a(\tilde x-\tilde y)}\,,
\ee
and then taking $a\rightarrow\infty$. The solution
(3.5) 
reduces to, for an appropriate choice of $K_0$,
\alpheqn\ba
\dif s^2&=&-\left(\frac{H(\tilde x)}{H(\tilde y)}\right)^{\frac{N}{3}}\dif t^2
+\frac{R^2}{(\tilde x-\tilde y)^2}\,\left(H(\tilde x)H(\tilde y)^2\right)^{\frac{N}{3}}\Bigg[
-\frac{1-\tilde y^2}{H(\tilde y)^N}\,\dif\psi^2\nonumber\\
&&\phantom{-\left(\frac{H(\tilde x)}{H(\tilde y)}\right)^{\frac{N}{3}}\dif t^2}-\frac{\dif\tilde y^2}{1-\tilde y^2}
+\frac{\dif\tilde x^2}{1-\tilde x^2}
+\frac{1-\tilde x^2}{H(\tilde x)^N}\,\dif\varphi^2\Bigg]\,,\\
A_\varphi&=&\sqrt{\frac{1+\mu}{1-\mu}}\,\frac{\sqrt{N}R\mu(1-\tilde x)}{H(\tilde x)}\,,\\
\phi&=&-\frac{\alpha N}{2}\ln\left(\frac{H(\tilde x)}{H(\tilde y)}\right),
\ea\reseteqn
where we have defined $H(\tilde\xi)=1-\mu\tilde\xi$ and 
$\mu=m/\sqrt{m^2+R^2}$. Indeed, this is precisely the solution for
the dipole ring, written in the coordinates of 
\cite{Emparan3}.\footnote{Alternatively, one could consider shrinking
the first black ring of Fig.~2 down to zero size, leaving just the second 
black ring. This limit also yields the solution for a single 
extremal dipole ring, and 
is in fact equivalent to taking the Myers--Perry black hole limit of 
the Figueras black ring \cite{Figueras}.}

We can also zoom in to each of the black rings directly. To zoom in
to the first one, we perform the coordinate transformation (\ref{bh1_x},c),
and then take $r$ to be smaller than any other length scale present. For an
appropriately chosen $K_0$, the resulting metric is 
\ba
\label{br1_metric}
\dif s^2&=&g(\theta)^{\frac{N}{3}}
\left[\left(\frac{r}{Q}\right)^{\frac{N}{3}}(-\dif t^2+r_1^2\dif\psi^2)
+\left(\frac{r}{Q}\right)^{-\frac{2N}{3}}(\dif r^2
+r^2\dif\theta^2)\right]\nonumber\\
&&\qquad
+\left(\frac{r}{Q}\,g(\theta)\right)^{-\frac{2N}{3}}
r^2\sin^2\theta\,\dif\varphi^2,
\ea
with the associated gauge field (\ref{bh1_gauge}) and dilaton 
(\ref{bh1_phi}). $Q$ and $g(\theta)$ continue to be given by (2.19). 
This solution describes the region near the
$S^1\times S^2$ horizon of an extremal dipole ring (with the horizon located at $r=0$). It can be seen
from the metric that the $S^1$ ring radius is 
$r_1\equiv[(1-r_+A)(1+r_-A)]^{\frac{1}{2}}/A$, while the $S^2$
sections of the horizon are distorted away from spherical 
symmetry by the 
presence of the factor $g(\theta)$. This distortion is expected from
the self-attraction of the ring, as well as the mutual attraction of the other
ring. Furthermore, it can be checked that this horizon is regular only 
in the non-dilatonic case $N=3$ (although it has zero area). For the dilatonic rings, the horizons
are actually null singularities.

Similarly, we may zoom in to the second ring with the transformation 
(\ref{bh2_x},c). The resulting solution is given by the same solution
as the above, but with the replacement $A\rightarrow-A$. In particular, 
the radius of this ring is $r_2\equiv[(1+r_+A)(1-r_-A)]^{\frac{1}{2}}/A$; 
as expected, it is greater than that of the first ring. Otherwise, the 
above remarks for the first ring still apply.

The distortion of the two ring horizons is an indication that there are
conical singularities attached to the rings, whose presence we shall now 
check for.
If we take the angles $\varphi$ and $\psi$ to have the usual periodicity 
$2\pi$, then the deficit angles along the three parts of the 
$\varphi$-axis are again given by (2.21), 
while the deficit angle along the $\psi$-axis is
\be
\delta_{(y=-1)}=2\pi\,\Big[1-(1-2mA-a^2A^2)^{\frac{N}{2}}\,K_0\Big]\,.
\ee
It can be seen that setting $K_0=(1-2mA-a^2A^2)^{-\frac{N}{2}}$
would simultaneously remove the conical singularities along 
the two semi-infinite axes $x=-1$ and $y=-1$. 
There will, however, be a resulting conical excess along $x=1$, and also along
$y=-\frac{1}{r_+A}$. The former is a disk spanning the interior of the 
first ring, whose presence is required to prevent the ring from collapsing under its
own attraction. The latter is an annulus spanning the region between
the two rings, keeping them apart in static equilibrium. 
Note that for this choice of $K_0$, the space-time is asymptotically
flat, and its ADM mass is readily computed to be $\frac{\pi}{2G}\frac{Nm}{A}$.

Now, it is possible to immerse the di-ring in a background magnetic 
field, using a five-dimensional analogue of the Harrison transformation 
\cite{Emparan2}. The resulting metric is 
\ba
\label{5Dmag_metric}
\dif s^2&=&-\left(\frac{H(y,x)\Lambda}{H(x,y)}\right)^{\frac{N}{3}}\dif t^2
+\frac{1}{A^2(x-y)^2}\,\left(H(y,x)H(x,y)^2\Lambda\right)^{\frac{N}{3}}\Bigg[
-\frac{(1-y^2)F(x)}{H(x,y)^N}\,\dif\psi^2\nonumber\\
&&+\frac{1}{K_0^2\,G(x,y)^{N-1}}\bigg(
-\frac{\dif y^2}{(1-y^2)F(y)}
+\frac{\dif x^2}{(1-x^2)F(x)}\bigg)
+\frac{(1-x^2)F(y)}{(H(y,x)\Lambda)^N}\,\dif\varphi^2\Bigg]\,,
\ea
with the associated gauge field and dilaton given by
(\ref{mag_gauge}) and (\ref{mag_phi}), respectively. In these
expressions, $\Lambda$ is defined as in (\ref{Lambda}), and
$B$ is a new parameter that determines the strength of the background 
magnetic field. This solution would no longer be asymptotically flat,
but would asymptote to a five-dimensional dilatonic analogue of the 
Melvin universe. 

It can be checked that the background magnetic field will contribute to the
distortion of the black-ring horizons. For the first black ring,
the near-horizon metric is given by (\ref{br1_metric}), with the 
distortion factor (\ref{g_mag}). For the second black ring, the
metric is the same but we have to replace $A\rightarrow-A$ in 
the distortion factor.
Furthermore, it can be verified that the presence of the
magnetic field would only affect the conical singularity along 
$y=-\frac{1}{r_+A}$, whose deficit angle is now given by (\ref{B_delta}).
By appropriately adjusting $B$, it is possible to remove this conical
singularity; in this case, there would no longer be any conical singularities 
attached to the second ring, and the $S^2$ sections of its horizon would 
become perfectly spherically symmetric. However, there will still be a 
conical excess along the inner disk $x=1$, which cannot be removed
by the magnetic field.

Finally, recall that in five dimensions, a solution that is magnetically
charged with respect to the gauge potential $A_a$ may be dualized into
another solution that is electrically charged with respect to a 
two-form gauge potential $B_{ab}$. The duality transformation takes
the form
\be
\phi'=-\phi\,,\qquad H_{abc}=\frac{\me^{-\alpha\phi}}{2}
\varepsilon_{abc}{}^{de}F_{de}\,,
\ee
where $H_{abc}\equiv\partial_aB_{bc}+\partial_bB_{ca}+\partial_cB_{ab}$, 
and maps the action (\ref{5D_action}) to
\be
\frac{1}{16\pi G}\int\dif^5x\,\sqrt{-g}\left(R-\frac{1}{2}(\partial\phi')^2
-\frac{1}{12}\me^{-\alpha\phi'}H^2\right).
\ee
Applying this duality transformation to the above solution would
give one that describes an electrically charged di-ring immersed in
a background electric field. The metric is still given by 
(\ref{5Dmag_metric}), while the dilaton is the negative of 
(\ref{mag_phi}). The two-form potential has a non-zero
component $B_{t\psi}$, which is given by the same expression as the
right-hand side of
(\ref{electric}). The properties of this solution would be
very similar to those of the magnetic solution.
For the particular case of $N=1$, it has the possible string-theory
interpretation as two loops of fundamental string with opposite
charges. More string-theory applications will be discussed in the
conclusion.

\newsection{Conclusion}

The accelerating dihole solution that we have presented in this
paper is, to the best of our knowledge, the first known example of 
a solution generalizing the charged C-metric to more than one 
black hole accelerating in the same direction.
(In the vacuum case, a solution describing multiple accelerating 
black holes has been constructed in \cite{Dowker3}.) While the 
existence of such a solution is perhaps not unexpected, what is
notable is the relatively compact form in which it can be written.
Unfortunately, this is unlikely to hold in any generalization of it.
For example, the non-extremal dihole solution is known to be
very complicated \cite{EmparanTeo}, and its accelerating 
version, if it can be found, would probably be even more so.

There are a variety of ways in which our solutions, 
particularly the di-ring, can be embedded in a higher-dimensional
theory such as string or M-theory. The simplest way is to extend the
world-volume of the di-ring by adding flat directions to it 
appropriately. One would then obtain concentric tubular brane--anti-brane
pairs. For example, one can use this method to 
construct tubular D6--anti-D6-branes, with world-volume geometry
$\mathbb{R}^{1,5}\times S^1$, in ten-dimensional Type IIA string theory.
Another way is to regard the di-ring as an intersection of higher-dimensional
branes. An M-theory realization of this consists of two sets of three 
M5-branes, with each set of M5-branes intersecting over a ring.
All these solutions contain conical singularities, which are necessary
to maintain equilibrium. Although it is possible to immerse them in a 
background magnetic field (also known as a fluxbrane in this context), 
this would not be sufficient to balance the forces present and remove 
all the conical singularities. For details
on the construction of these solutions, we refer the reader to
\cite{Emparan2,Chattaraputi}.

Finally, we note that the rotating black ring solution of Figueras
\cite{Figueras} can be turned into yet another solution of five-dimensional
vacuum Einstein gravity. This solution can be read off from the results
of Sec.~2, by taking the electrically charged accelerating dihole 
solution in Kaluza--Klein theory, and lifting it back to five dimensions
with the appropriate analytic continuations.
It can be checked that this new solution describes a pair of 
concentric, extremally rotating black rings. The rotation of these 
rings is now in the $S^1$ direction, as in the original rotating 
black ring of Emparan and Reall \cite{EmparanReall2}. Moreover, 
they are rotating in directions {\it opposite\/} to each other. 
Because their rotation is extremal (in the sense that their 
corresponding rods have shrunk down to zero length), 
these rings are actually null singularities. Nevertheless, 
it should be interesting to study this solution in more detail.

\appendix

\newsection{Accelerating Zipoy--Voorhees solution}

Of the black dihole class of solutions, the one belonging to pure 
Einstein--Maxwell theory is perhaps the most famous. It was first
derived by Bonnor \cite{Bonnor} in 1966, who recognized that it
represented a magnetic dipole source; for this reason, it is also known as the
Bonnor dipole solution in the literature. In his paper, Bonnor noted 
that taking the coincident limit of this solution yields the neutral 
Darmois solution \cite{Darmois}, 
which can be interpreted as two Schwarzschild black holes superposed on the
top of each other in the Weyl formalism. The Darmois solution is, in fact,
a special case of a more general class of static, axisymmetric 
vacuum solutions known as
the Zipoy--Voorhees solution \cite{Zipoy,Voorhees}. The latter
is parametrized by the real number $\delta$, and 
can be interpreted as the
superposition of $\delta$ Schwarzschild black holes when $\delta$ is 
a positive integer. The usual 
Schwarzschild solution is recovered when $\delta=1$. For all other
values of $\delta$ however, the space-times contain naked singularities,
and so their physical interpretations remain unclear.

It therefore follows that the coincident limit of the accelerating Bonnor
dipole solution is an accelerating version of the Darmois solution. It
can be read off from (\ref{metric}), by setting the dilaton coupling
$\alpha$ and separation parameter $a$ to zero.
In fact, we can do better: it is possible to generalize this solution
to include an arbitrary parameter $\delta$, thereby obtaining an 
accelerating version of the Zipoy--Voorhees solution. The result is
\ba
\label{accel_VZ}
\dif s^2&=&\frac{1}{A^2(x-y)^2}\Bigg[
\left(\frac{F(y)}{F(x)}\right)^{\delta}
(1-y^2)F(x)\,\dif t^2+\left(\frac{F(y)}{F(x)}\right)^{-\delta}(1-x^2)F(y)\,\dif\varphi^2\nonumber\\
&&\phantom{\frac{1}{A^2(x-y)^2}\Bigg[}+\frac{(F(x)F(y))^{\delta(\delta-1)}}{G(x,y)^{\delta^2-1}}\,\Bigg(-\frac{\dif y^2}{(1-y^2)F(y)}
+\frac{\dif x^2}{(1-x^2)F(x)}\Bigg)\Bigg]\,,
\ea
where 
\be
F(\xi)=1+2mA\xi\,,\qquad
G(x,y)=(1+mA(x+y))^2-m^2A^2(1-xy)^2.
\ee
Note that the usual C-metric (in the form written in \cite{Hong}) is 
obtained when $\delta=1$, while the accelerating Darmois solution is 
obtained when $\delta=2$.
If we perform the coordinate transformation (\ref{zero_acc}) and take 
the zero acceleration limit $A\rightarrow0$, we indeed recover the 
Zipoy--Voorhees solution from (\ref{accel_VZ}):
\ba
\dif s^2&=&-\left(1-\frac{2m}{r}\right)^{\delta}\dif t^2
+\frac{r^{\delta(\delta+1)}(r-2m)^{\delta(\delta-1)}}{[(r-m)^2-m^2\cos^2\theta]^{\delta^2-1}}\,\Bigg(\frac{\dif r^2}{r(r-2m)}
+\dif\theta^2\Bigg)\nonumber\\
&&\qquad+\left(1-\frac{2m}{r}\right)^{1-\delta}r^2\sin^2\theta\,\dif\varphi^2.
\ea
It should be possible to analyze the space-time (\ref{accel_VZ}) in more
detail using standard methods, but we will not do so here.

\bigskip\bigskip

{\renewcommand{\Large}{\normalsize}
}
\end{document}